\documentclass[a4paper,3p,10pt,times]{elsarticle}
\usepackage{graphicx}
\usepackage{epsfig}
\usepackage{amssymb}
\usepackage{natbib}
\usepackage{hyperref}
\usepackage{amsmath}
\usepackage{color}
\usepackage{verbatim}
\usepackage[usenames,dvipsnames]{xcolor}
\usepackage{pstricks,pst-node,pst-text,pst-3d}
\usepackage{verbatim}
\usepackage{ulem}

\newcommand{\tn}{\textnormal}
\newcommand{\ve}{\varepsilon}
\newcommand{\defeql}{\ \mathrel{\mathop:}=\ }
\newcommand{\mh}{\mathbf}
\newcommand{\mc}{\mathcal}
\journal{Physica A}
\begin{document}
\begin{frontmatter}

\title{Extensive numerical investigations on the ergodic properties \\
  of two coupled Pomeau-Manneville maps}
\author[mpiks]{Matteo Sala}
\ead{msala@pks.mpg.de}
\author[udesc,uninsubria]{Cesar Manchein}
\ead{cesar.manchein@udesc.br}
\author[uninsubria,milano]{Roberto Artuso}
\ead{roberto.artuso@uninsubria.it}
\address[mpiks]{Max Planck Institute for the Physics of Complex
  Systems, N\"{o}thnizer Stra{\ss}e 38, 01187 Dresden (Germany);} 

\address[udesc]{Departamento de F\'\i sica, Universidade do Estado de
  Santa Catarina, 89219-710 Joinville, (Brazil);}

\address[uninsubria]{Center for Nonlinear and Complex Systems and
  Dipartimento di Scienza e Alta Tecnologia, Via Valleggio 11, 22100
  Como (Italy);} 

\address[milano]{I.N.F.N., Sezione di Milano, Via Celoria 16,
  20133 Milano (Italy)}  

\begin{abstract}
We present extensive numerical investigations on the
ergodic properties of two identical Pomeau-Manneville maps interacting
on the unit square through a diffusive linear coupling. The system
exhibits anomalous statistics, as expected, but with strong deviations
from the single intermittent map: Such differences are characterized
by numerical experiments with 
densities which {\it do not} have singularities in the marginal fixed
point, escape and Poincar\'e recurrence time statistics that share 
a power-law decay exponent modified by a clear {\it dimensional}
scaling, while the rate of phase-space filling and the convergence of
ensembles of Lyapunov exponents show a {\it stretched} instead of
pure exponential behaviour. In spite of the lack of rigorous results
about this system, the dependence on both the intermittency and the
coupling parameters appears to be smooth, paving the way for further
analytical development. We remark that dynamical exponents appear to
be independent of the (nonzero) coupling strength. 
\end{abstract}
\begin{keyword}
Intermittency \sep coupled maps \sep Poincar\'e recurrences \sep
escape times \sep finite-time Lyapunov exponents. 
\end{keyword}
\end{frontmatter}
%
\section{Introduction}
\label{Introduction}
Intermittent systems represent a paradigmatic example of {\it weak
  chaos}, where chaotic domains coexists with regular, non-chaotic
structures. Typical examples of such a situation involve area
preserving maps (see for example \cite{lic,zas,ces}), where regular
structures often occupy a finite fraction of the whole phase-space. On
the other side, intermittent systems having a {\it zero-measure}
source of non-hyperbolicity
or regular motion often allow more quantitative considerations,  a
typical example being Pomeau-Manneville (PM) maps~\cite{PM}, 
for which many peculiar properties (from power-law correlation decay to
generalised central limit theorems to infinite ergodic theory), have
been anticipated by Gaspard and Wang~\cite{GW,Wang} and later
discussed and proved in several papers, with also remarkable
applications to the problem of anomalous transport. 

In this work we address the important problem of multidimensional
extension of such prototypical models by {\it diffusively}
coupling~\cite{GrPi,fer} a 
pair of Pomeau-Manneville maps: it is not trivial, indeed, how the
increased dimensionality may modify the dynamical properties, {\it e.g.}
allowing the system to avoid or deform the intermittent behaviour.
In other words, we are interested in characterising the ergodic
properties of two coupled PM maps and compare this case with the
single PM map results. Among the different tools to investigate such
properties a prominent role is played by the analysis of 
phase-space densities, escape and recurrence time statistics, filling
rates and finite-time Lyapunov statistics. Apart from the
intrinsic interest, this study represents a natural preliminary step
towards many-dimensional extensions, which have been recently
proposed, for example, as the proper framework to study the statistics
of genomic sequences~\cite{ast}.  

The paper is organised as follows: Section~\ref{fr} describes general
properties of the 1D PM map and its 2D version (PM2)
coupled via the diffusive scheme; stability arguments allow to
identify and avoid the so-called synchronization regime. In
Section~\ref{nr} we present and discuss the numerical results that are
summarised in the conclusions in Section~\ref{sec:conc}. 

\section{General setting}
\label{fr}
\subsection{Pomeau-Manneville (PM) map}
\label{pmm}
\begin{figure}[!htb]
  \centering
  \includegraphics[width=\textwidth]{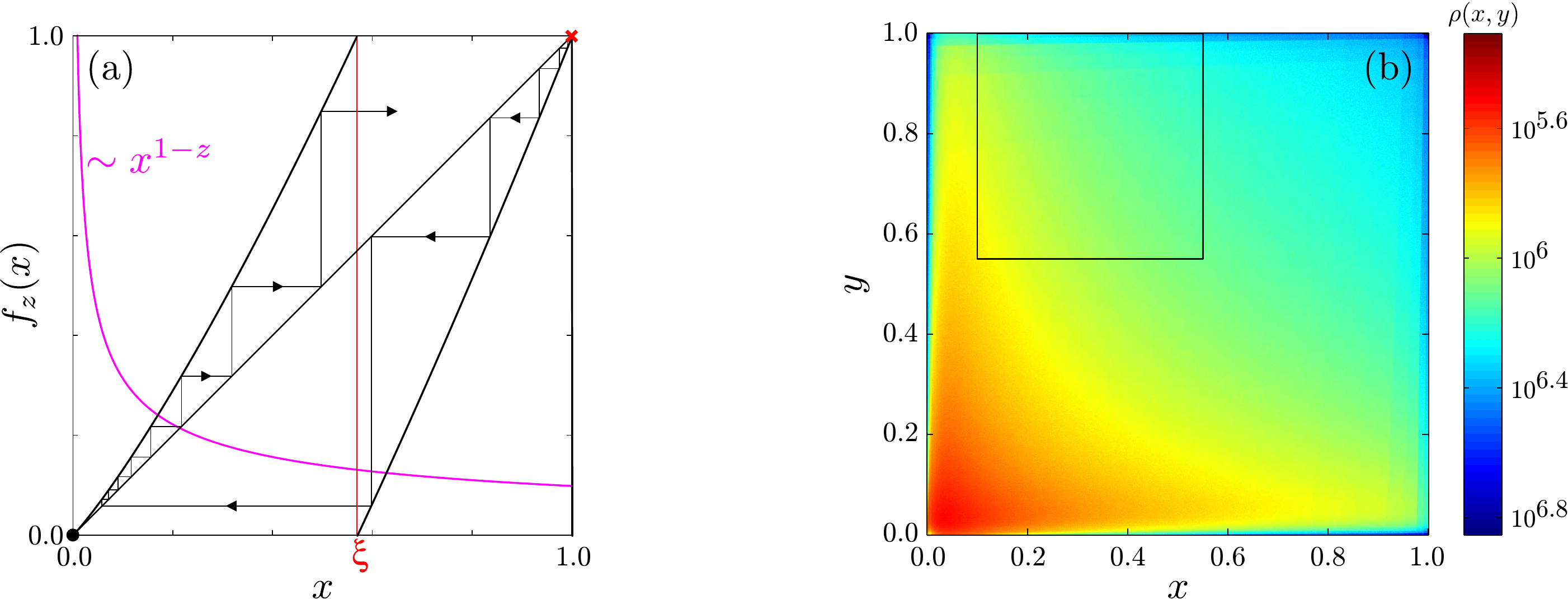}
  \caption{(Color online) Panel (a): Phase-space dynamics of the PM
    map, given by Eq.~(\ref{pm1}), {with $z=1.5$ and
      $\xi(z)\simeq0.57$},  along with a typical orbit: after being
    quickly repelled away from the unstable fixed point at $x=1$ (red
    cross) and being re-injected into $[0,\xi]$, it slowly escapes
    from the indifferent fixed point at $x=0$ (black point). The
    nearer the reinjection is to $x=0$, the longer the escape lasts,
    leading to an invariant density with a singularity $\sim x^{1-z}$
    at $x=0$ (magenta-continuous line). Panel (b): {in color, the 
      two-dimensional numerical density of points $\rho(x,y)$ (the
      color-bar is presented in logarithmic scale)} for an orbit of
    length $10^{11}$ of the coupled PM2 map (see Eq.~\ref{2d}) with
    $z=1.5$ and $\ve=10^{-2}$; the orbit visits the whole partition of
    $10^3\times10^3$ cells, but the color levels reveal that the
    density is not the product of two PM densities. The upper left
    square is the region over which recurrences are computed
    (Sec.~\ref{sec:rec}); this is chosen to avoid the synchronized
    states on the diagonal.} 
  \label{fig:1}
\end{figure}

\noindent
The PM map is a non-invertible transformation of the unit-interval
$f_z:[0,1]\circlearrowleft$, defined by: 
\begin{align}
	x\ \mapsto\ \nonumber f_z(x)\ &=\ x\,+\,x^z\mod 1\\
	&=\ x\,+\,x^z\,-\,\chi_z(x),
	\label{pm1}
\end{align}
where $1<z\in\mathbb{R}$ and $\chi_z(x)$ is the \textsl{characteristic
function} of the set $[\xi,1]$ {(\textit{i.e.} $\chi_z(x)=0$
for $x<\xi$, and $\chi_z(x)=1$ for $x \geq \xi$)} with the
discontinuity point $\xi(z)<1$ defined by $f_z(\xi)=1$. {As we
  remarked in the introduction, we are interested in the regime where
  the fixed point 
at $0$ is not hyperbolic: this happens whenever $z>1$; notice that
when $z=1$ (\ref{pm1}) yields the -fully chaotic- Bernoulli
map). Thus, for $z>1$, at the fixed point at the origin we have
$f'(0)=1$, namely we have an \textit{indifferent} fixed point,
neither stable nor unstable, as far as linear analysis is
concerned.} The dynamical properties of the PM 
map are determined by its indifferent fixed point at $x=0$, where the
dynamics is slowed down by the tangency {(see
  Fig.~\ref{fig:1}(a))}, and 
they strongly depend on the \textsl{intermittency parameter} $z$. 
When $z>2$ there is no invariant probability measure (the invariant
density close to the origin goes as $x^{1-z}$) and the map (\ref{pm1})
provides and example of {\it infinite ergodicity}~\cite{aa,Zw1,BB}: We
will not consider such a case in the present paper. On the other hand,
we are interested in the regime $1<z<2$ where the map is ergodic and
the invariant measure close to the origin has the same behaviour as
before, but now with an integrable singularity~\cite{po}. For
such intermittency parameter range the map is also  
mixing, with polynomial decay of correlation functions: the
corresponding power-law exponent $\Gamma$ (such that, asymptotically,
the mixing speed is $n^{-\Gamma}$) can be expressed in terms of the
intermittency parameter as follows~\cite{Wang,Hu}:  
\begin{equation}
\label{1Ecorr}
\Gamma=\frac 1{z-1}-1.
\end{equation}
Notice that, when $z \geq 3/2$, correlations are not integrable
{\textit{i.e.} $\int\,dt\,C(t)$ diverges}: a striking dynamical
manifestation of this observation is that {a generalized central
  limit theorem holds with properly scaled sums converging to } a
L\'evy stable law~\cite{Gou}.  

\subsection{Coupling scheme}
We start our analysis by considering two {\it identical} copies of
the PM map $f_z$ given by Eq.~(\ref{pm1}) interacting on the unit
square $S=[0,1]^2$ through the application of a linear {\it diffusive}
coupling; this leads to the following 2D (PM2) map:  
\begin{align}
  \mh{f}_{z,\ve}:\ \left[
\begin{array}{c}
  x\\
  y
\end{array}
\right]\ \mapsto\ 
\left[
  \begin{array}{c}
    f_z\left(\,x\,+\,\ve\left(y-x\right)\,\right)\\
    f_z\left(\,y\,-\,\ve\left(y-x\right)\,\right)
  \end{array}
\right]\ ;
\label{2d}
\end{align}
which corresponds to the map considered in Ref.~\cite{GrPi} with
the nonlinear and coupling parts applied in {\it reverse order}.
This choice allows to define Poincar\'e recurrences in a proper way:
indeed, the coupling part maps the unit square $S$ into a smaller sub-set
contained in S (see Fig.~1 in Ref.~\cite{GrPi}) which then becomes the
actual phase-space for the map. Instead, by applying the nonlinear part
{\it after} the coupling, as in Eq.~(\ref{2d}), we ensure that the
mapping is {\it onto} $S$. 
The linear coupling in Eq.~(\ref{2d}) is the two-maps
limit of the renown {\it Laplacian} or {\it diffusive} coupling
for a chain of $N$ coupled maps $\{x_j\}_{j=1..N}$ with periodic
boundary conditions $x_{N+1}=x_1$~\cite{fer}. The choice of
this coupling is motivated by its simplicity, being linear, and by the
fact that it converges to the Laplace operator in the limit
$N\to\infty$ of a {continuous} {closed} chain of maps.

\subsubsection{Linear stability}
In the case of two coupled maps, the {\it diagonal} $x=y$ of the unit
square is an {\it invariant set} for map (\ref{2d}) (we may call it
the set of {\it synchronized} states of the
system); it is thus informative to study its stability properties. For
points $\mh{x}=(x,x)$, the first/second eigenvector of the Jacobian
matrix of map (\ref{2d}) is respectively parallel/orthogonal 
to the diagonal itself; a calculation analogous to Ref. \cite{GrPi}
allows to compute the two Lyapunov exponents of the diagonal: 
\begin{align}
  \lambda_1^{diag} = \lambda^{1\tn{D}}, \quad
  \lambda_2^{diag} = \lambda^{1\tn{D}}+\ln(1-2\ve) < \lambda_1; 
\end{align}
with $\lambda^{1\tn{D}}$ the Lyapunov exponent of the 1D
Pomeau-Manneville map (\ref{pm1}) with initial condition $x$; from the
expression for the second exponent $\lambda_2^{diag}$ one can see that, when
the coupling parameter $\ve$ is large enough, $\lambda_2^{diag}$ becomes negative
and the diagonal turns into an {\it attracting set}, leading to global
{synchronization}~\cite{syn1,syn2} (that is, the system {synchronizes}
for any initial condition in the unit square $S$). Thus, by setting
$\lambda_2^{diag}=0$, the critical coupling value $\ve_{cr}$ above
which {synchronization} appears is obtained:   
\begin{align}
\ve_{cr} = \frac{1-e^{-\lambda^{1\tn{D}}}}{2},
\end{align}
and it depends on parameter $z$ trough the exponent $\lambda^{1\tn{D}}$;
this relation holds for any pair of diffusively coupled,
{\it identical} 1D maps. In the following we will only consider the
range $\ve \ll \ve_{cr}$, {\it i.e.}, avoiding {synchronization}; this
is achieved by a preliminary estimate of the 1D Lyapunov exponent 
$\lambda^{1\tn{D}}$ for the chosen value of $z$. Notice that, in the
range $z\in]1,2[$ we consider here, the Lyapunov exponent 
$\lambda^{1\tn{D}}$ is always positive for any initial condition
$\mh{x}=(x,x)\neq(0,0)$; instead, the exponent $\lambda^{1\tn{D}}$
{\it vanishes} for the fixed point in the origin $(0,0)$, so the two
Lyapunov exponents for such point read: 
\begin{align}
  \lambda_1^{(0,0)} = 0,	\quad\lambda_2^{(0,0)} = \ln(1-2\ve) < 0.
\end{align}
This implies that the stability of fixed point $(0,0)$ is 
{\it discontinuous} in $\ve=0$ : for $\ve>0$ it is {\it marginal}
along the diagonal ($\lambda_1^{(0,0)}=0$) and {\it contracting}
along the anti-diagonal ($\lambda_2^{(0,0)}<0$), while for $\ve=0$ it
is marginal along both the eigenvectors, since
$\lambda_2^{(0,0)}\to\lambda_1^{(0,0)}=0$ smoothly as
$\ve\to0$. Interestingly, this scenario holds even in the 
{non-synchronizing} regime, when the diagonal is repelling
($\lambda_2^{diag}>0$).  
This is a first hint about the {\it structural} change in the map
caused by the coupling: in the uncoupled case $\ve=0$, the whole
neighbourhood of the origin is affected by its marginal stability and
the associated slow-down; instead, in the coupled case $\ve\neq0$ the
slow-down affects {\it only} the diagonal, while the infinitesimal
neighbourhood of the origin is {\it contracted} toward the diagonal
along its transversal direction. This is confirmed by the numerical
phase-space densities presented in the next Section. 

\section{Numerical results}
\label{nr}
\subsection{Ergodicity test 1: Phase-space densities}
\label{sec:dens}
\begin{figure}[!htb]
  \centering
  \includegraphics[width=\textwidth]{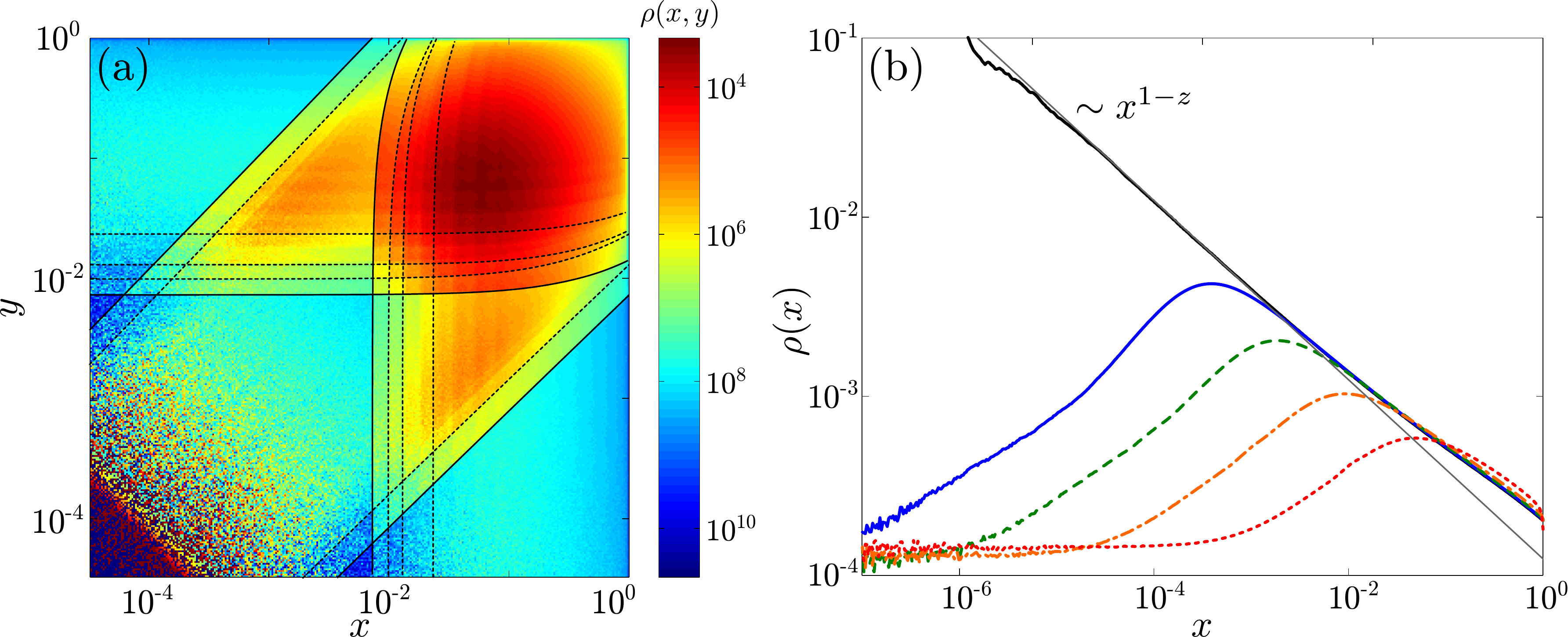}
  \caption{(Color online) Panel (a): in color, the
    two-dimensional numerical density of points $\rho(x,y)$ (the
    color-bar is presented in logarithmic scale) for an orbit of 
  length $10^{11}$ collected  over a logarithmic square-grid of
  $10^3\times10^3$ for the PM2 map (\ref{2d}) with $z=1.5$ and
  $\ve=10^{-2}$. The black-continuous (-dashed) curves are
    respectively, the first (second) iterates of the axes $x=0$ and
    $y=0$. Over such curves and also over the higher iterates  (not
    shown), the density appears discontinuous. Panel (b): marginal 
  distributions $\rho(x)$ for the coordinate $x$ of the same orbit
  in panel (a) for $\ve=0$ (black curve), $10^{-5}$ (blue curve),
  $10^{-4}$ (green-dashed curve), $10^{-3}$ (orange-dash-dotted
  curve) and $10^{-2}$ (red-dotted curve); the gray lines is the
  standard estimate $\sim x^{1-z}$ for the PM density.}    
  \label{fig:2}
\end{figure}

To analyse the phase-space structure of the PM2 map, we
compute numerical densities for orbits of length $10^9$ over a square
grid of $10^3\times 10^3$ cells {(Fig.~\ref{fig:1}(b))}: this shows
that no dynamical barriers are present, with the orbit visiting all
the cells of the partition even for arbitrarily small
  couplings. The observation of inhomogeneities in a small  
neighbourhood of the origin leads us to perform the same phase-space
analysis employing a square-grid partition in {\it logarithmic} scale for
coordinates $x$ and $y$, as presented in the {Fig.~\ref{fig:2}(a)}:
this reveals that the density is dominated by a very complicated
structure of {\it discontinuity lines} which delimit regions with much
higher density. The reading of such 
logarithmic color-map should be compared with its corresponding linear
version in Fig.~\ref{fig:1}(b), to conclude that the observed orbit
tends to avoid the coordinate axes while visiting much more frequently
the central maximum away from the origin. Such behaviour is in clear opposition
to the classic divergence of the PM density in the origin (see also
Fig.~\ref{fig:1}(a)) and is reflected in the {\it marginal} distributions
for the single coordinate $x$ for the PM2 map {(Fig.~\ref{fig:2}(b))}:
a comparison with the corresponding distribution for a PM orbit with
the same intermittency parameter shows that, while the 1D distribution
diverges in the origin as $\sim x^{1-z}$, the 2D distributions for all $\ve\neq0$ 
have a maximum away from the origin; this appears to move continuously
toward $x=0$ as $\ve\to0$. This phenomenon is robust against different
sizes of phase-space partition, proving that the effect of {\it arbitrarily}
small couplings have a strong impact on both the dynamics and statistics 
of the system. Even more interestingly, the location of the discontinuity
lines in the PM2 phase-space densities can be computed {\it analytically}:
in Fig.~\ref{fig:2}(a), the solid and dashed curves are respectively
the first and second iterates of the coordinate axes $x=0$ and $y=0$,
while all the remaining discontinuities are checked to be higher iterates
of the same sets. This unexpected phenomenon is in agreement with the
estimates of the Lyapunov exponents from previous Section: the fixed
point in the origin $(0,0)$ is no more completely marginal, but has one
negative Lyapunov exponent which {\it attracts} the orbit toward the
diagonal; at the same time, we choose parameters for which the
diagonal is instead {\it repelling}. This produces the net effect of
enhancing the density in a region away from the origin, differently
from the uncoupled and 1D case.

\subsection{Ergodicity test 2: Filling rate}
\label{sec:fill}
\begin{figure}[!htb]
  \centering
  \includegraphics[width=\textwidth]{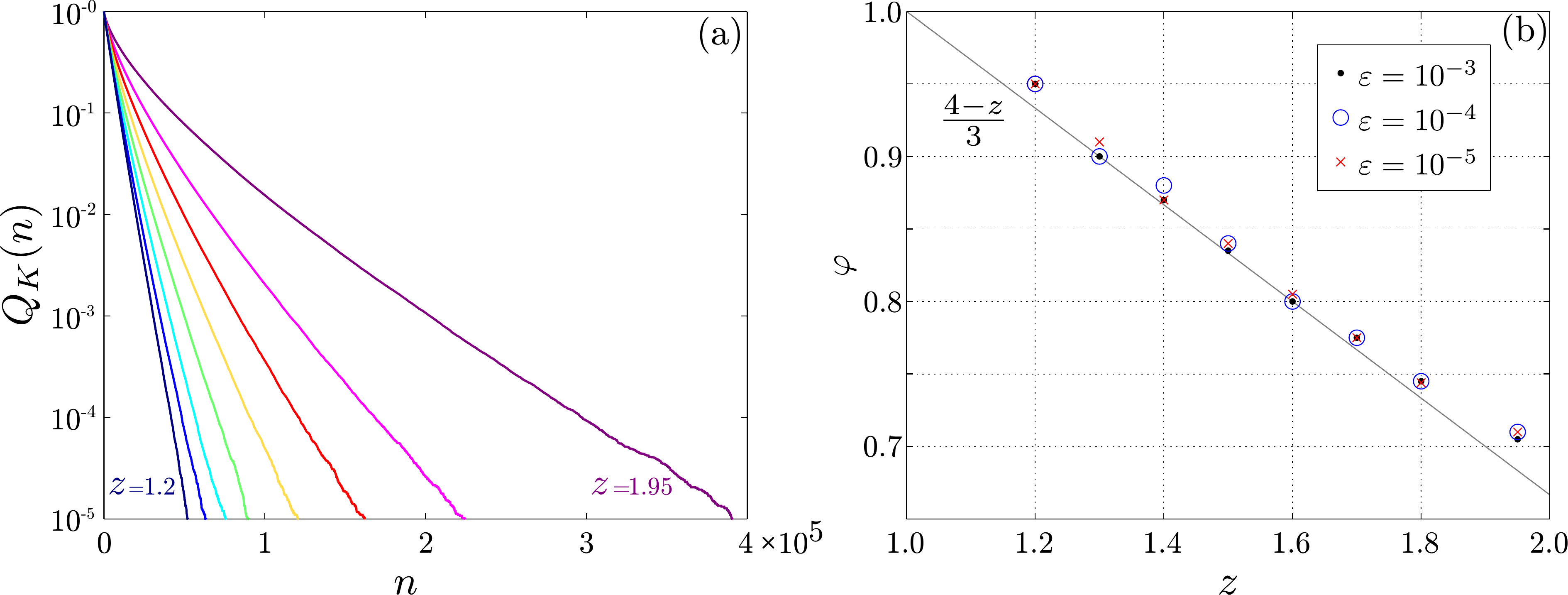}
  \caption{(Color online) Panel (a): stretched exponential decay of
    $\mc{Q}_K(n)$ 
    (fraction of unvisited cells at time $n$) for the values of
    intermittency exponent $z=1.2,1.3,1.4,1.5,1.6,1.7,1.8,1.95$
    (respectively from left- to right-most graph) at 
    coupling parameter $\ve=10^{-4}$; the total number of cells
    $K=2000\times2000$ is eventually visited, but the plot is
    truncated at $\mc{Q}_K(n)=10^{-5}$ for clarity. Panel (b): the
    stretching-exponent $\varphi$ from Eq.~(\ref{stQ}) as a
    function of $z$ for the three coupling parameters $\ve=10^{-3},\
    10^{-4},\ 10^{-5}$; in all three cases the almost linear trends
    are close to the line $(4-z)/3$ (in grey, for reference), leading
    to pure exponential at $z=1$ ({\it i.e.} the Bernoulli map
    limit).} 
  \label{fig:3}
\end{figure}

To characterize the ergodic properties of the PM2 map we employ the
{\it filling rate} method discussed in Ref.~\cite{rob} for the same
phase-space partition of $K=2000\times 2000$ square cells, and
consider the following indicator: 
\begin{equation}
  \label{fillF}
  \mc{Q}_K(n)=\frac{\textit{number of unvisited cells up to the n-th
      iteration}}{K}.
\end{equation}
which represents the fraction of partition that is not yet
visited by an orbit after $n$ iterations. A numerical verification of
ergodicity is equivalent to 
check that function $\mc{Q}_K(n)$ vanishes in the long time limit
$n\to\infty$, independently on how fine we choose the phase-space
partition ({\it i.e.} in the large $K$ limit). In Fig.~\ref{fig:3}(a),
a set of examples shows how indeed $\mc{Q}_K(n)$ converges to zero,
confirming the observations from previous section (eventually, the orbit
visits all the cells of the grid for all values of parameters $\ve$ and
$z$ considered here, see Fig.~\ref{fig:3}(b)). In addition to the
qualitative check for ergodicity, it is informative to study also how
$\mc{Q}_K(n)$ decays to zero: for models
of fully chaotic dynamics, this is expected to be a pure exponential
~\cite{rob} $\mc{Q}_K(n)\sim e^{-n/K}$. While generalisations of such
a decay have been proposed~\cite{rob2}, we remark that in \cite{cpmap}
a pure exponential decay has been observed, despite the model under
investigation is a billiard table shaped like a fully irrational
triangle, a system with very weak ergodic properties.
In our simulations we never get any evidence of such a simple asymptotics,
while the data are fitted by {\it stretched exponentials}, with stretching
exponent $\varphi$ and pre-factor $C$: 
\begin{equation}
\label{stQ}
\mc{Q}_K(n)\sim e^{-C\,\cdot\,n^{\varphi}}.
\end{equation}
The resulting exponents are plotted in Fig.~\ref{fig:3}(b), which suggests
a linear relation with the intermittency parameter
$z$ that is {\it independent} on the coupling parameter $\ve$;
interestingly, the latter turns out to influence only the exponent
pre-factor $C$ (not shown), a property shared also by other dynamical
indicators we consider in the next sections. Stretched exponential
relaxations are known to have physical relevance (see for example
Refs.~\cite{phil,DS}): in the present context we have not any
dynamical clue for this behaviour, which will be observed for other
dynamical quantities in the final part of the present section. 


\subsection{A warm up exercise: Escape from the indifferent region}
\label{sec:esc} 
\begin{figure}[!htb]
  \centering
  \includegraphics[width=0.6\linewidth]{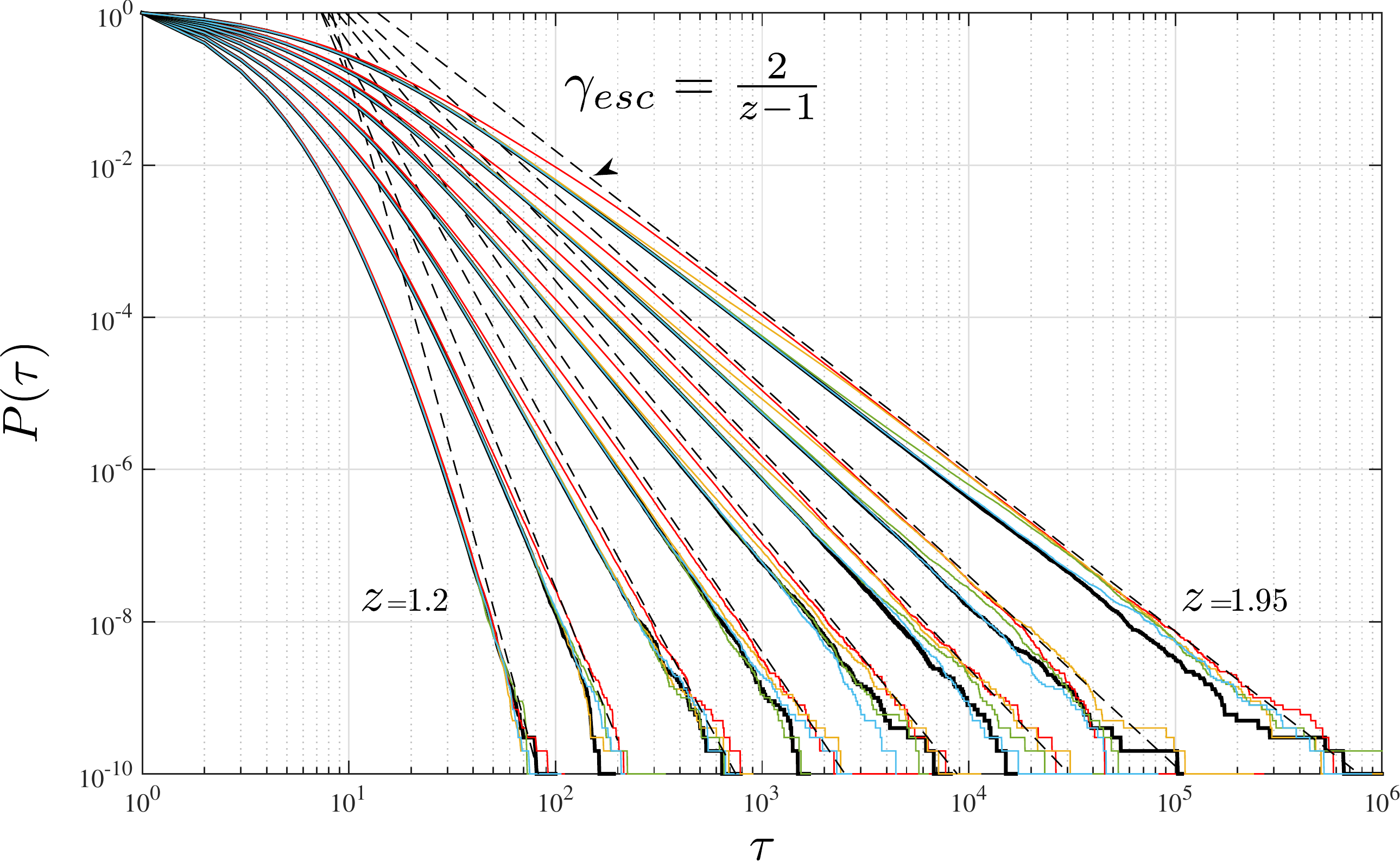} 
  \caption{(Color online) Cumulative probability distributions of escape times for 2
    coupled PM maps with $z=1.2,1.3,1.4,1.5,1.6,1.7,1.8,1.95$
    and $\ve=10^{-2}$ (red-continuous curves), $\ve=10^{-3}$
    (yellow-continuous curves), $\ve=10^{-4}$
    (green-continuous curves), $\ve=10^{-5}$
    (blue-continuous curves) 
    and $\ve=0$ (black-continuous curves); the escape is computed
    from the hyper-cube with side $0.1$ located in the corner of the
    origin. All numerical data agree very well with the estimate
    $\gamma_{esc}=2/(z-1)$ represented by dashed-black curves.}  
  \label{fig:4}
\end{figure}

It is well known that direct numerical indications of mixing are
extremely hard to attain for non trivial systems~\cite{PDcorr}: tests 
that proved to be numerically stabler involve either statistics of
Poincar\'e recurrences~\cite{Leb,Dim,bill,fam} or large deviations for
Birkhoff sums~\cite{am}. 
As a preliminary step, we consider escape rates from a small region of
the phase space containing the origin: for intermittent dynamics we
expect a power-law decay of the surviving probability. Our numerical
results refer to the choice of the escape region
$\Omega=\{(x,y)\in[0,0.1]^2\}$, we checked that our results remain
unaltered for smaller escape sets. The plots presented in
Fig.~\ref{fig:4} are obtained by choosing a uniformly distributed
ensemble of $10^{10}$ initial conditions in $\Omega$, and computing
the cumulative survival probability $P_{\Omega}(\tau)$. For uncoupled
($\ve=0$) maps, since the set of points that survive $n$ iterations is
a square, whose construction is easily induced by the 1D
case~\cite{Wang}, we expect that $P_{\Omega}(\tau)\sim
\tau^{-\gamma^{\ve=0}_{esc}}$ where the exponent is twice the value
for 1D PM maps, $\gamma^{\ve=0}_{esc}=2/(z-1)$. From Fig.~\ref{fig:4}
we observe that the escape power laws follow very closely the
uncoupled behaviour over a wide set of coupling constants $\ve$. This
result is however only a probe of local properties close to the
indifferent fixed point, so we now turn to more meaningful tests of
mixing properties. 

\subsection{Mixing test 1: Recurrence times}
\label{sec:rec}
\begin{figure}[!htb]
  \centering
  \includegraphics[width=\textwidth]{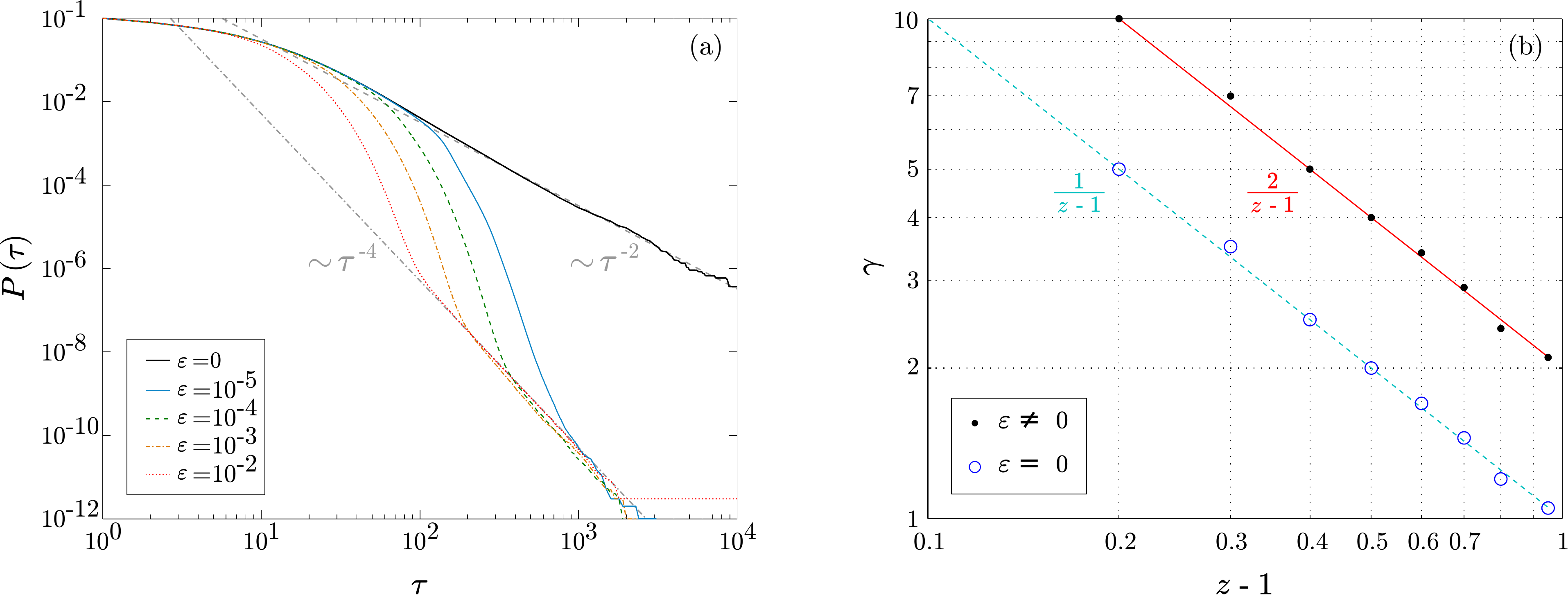}
  \caption{(Color online) Panel (a): Cumulative probability for
    recurrence times $\tau$ at $z=1.5$ and coupling
    parameter $\ve=10^{-2},10^{-3},10^{-4},10^{-5},0$; dashed and
    dash-dotted lines are 
    fits to the power-law decay of respectively the uncoupled
    ($\ve=0$) and coupled ($\ve>0$) cases. Panel (b): The same
    power-law exponents for the set of intermittency exponents under
    study $z=1.2,1.3,1.4,1.5,1.6,1.7,1.8,1.95$; we find the same exponents
    for all $\ve>0$ cases (black points, red-continuous line $\sim  
    2/(z-1)$ for reference), and all of them coincide with
    {\it twice} the $\ve=0$ exponents (blue circles, 
    light-blue-dashed line $\sim 1/(z-1)$ for reference). Notice that
    the characteristic time at which the power-law starts
    grows as the coupling parameter $\ve\to0$.}
  \label{fig:5}
\end{figure}

\begin{figure}[!htb]
  \centering
  \includegraphics[width=\textwidth]{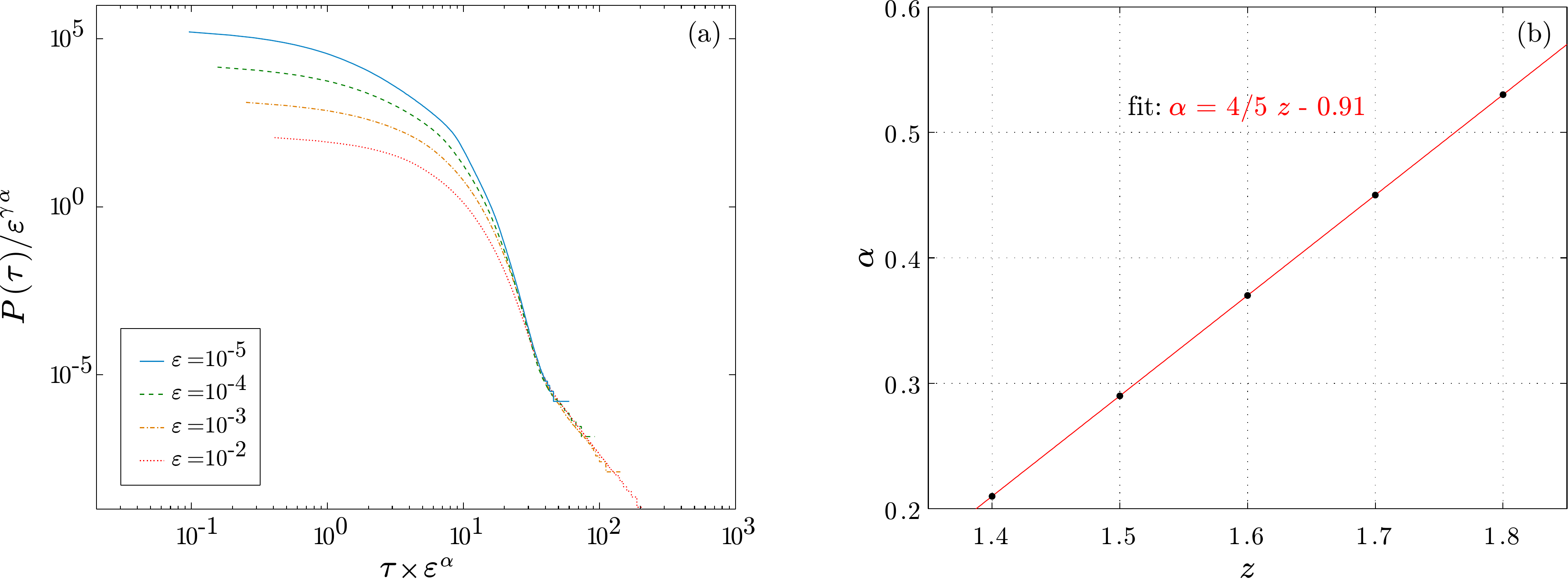}
  \caption{(Color online) Panel (a): Cumulative probability of recurrence times
    $\tau$ at $z=1.5$ (same as in Fig.~\ref{fig:5}(a)) after rescaling
    according to Eq.~(\ref{Pscal}). Panel (b): The rescaling
    exponent $\alpha$ as a function of parameter $z$, for the cases in
    which the curve collapse is apparent; the solid-red line is a
    linear fit.} 
  \label{fig:6}
\end{figure}

In this Subsection we present numerical
experiments for the statistics of Poincar\'e recurrences in a fixed
box in phase space: we choose
$(x,y)\in[0.10,0.55]\times[0.55,1.00]={\cal X}\times{\cal Y}$ away
from the diagonal (synchronized states), and a single random initial 
condition was iterated until the ensemble of recurrences was equal to
$10^{12}$. We also tested boxes with different sizes and  the
quantitative results remain unaltered. We remark that such a huge
ensemble of recurrences is essential to get a reliable statistics for
long return times. To visualize the data we plot the
cumulative probability $P_{rec}(\tau)\defeql P(t_{rec}\geq\tau)$ for
the orbit to return in the box only {\it after} a certain time
$\tau$: if the cumulative 
probability decays polynomially with an exponent $\gamma$, then we
expect~\cite{Leb,Dim,bill} that also correlations display a power-law 
decay, with an exponent $\Gamma=\gamma-1$. By such analysis, in 
Fig.~\ref{fig:5}(a) we confirm the presence of an asymptotic power-law
decay:  
\begin{equation}
P_{rec}(\tau)\sim\tau^{-\gamma}
\end{equation}
whose exponent $\gamma$ depends on the intermittency parameter $z$ but
not on the diffusive coupling $\ve$. As illustrated in
Fig.~\ref{fig:5}(b), such dependence for 
$\ve > 0$ coincides with {\it twice} the exponent behaviour for the
case $\ve=0$ , that is $\gamma(z)\sim1/(z-1)$, namely the exponent of
the 1D map (see Ref.~\cite{Gei} for an early derivation of such a
formula, based on escape rates from the indifferent fixed point). 
We may understand the $\ve=0$ behaviour by a qualitative argument as
follows: if we consider return times on a {\it strip} 
$(x,y)\in {\cal X}\times [0,1]$ then the statistics is 
determined by the single PM map for the variable $x$, and so
Poincar\'e recurrences $n_1,\,n_2\,\cdots n_k\cdots $ are ruled by the
exponent of the 1D 
map. If we now slice the strip into $M$ pieces, one being ${\cal
  X}\times{\cal Y}$, returns are now $\tilde{n}_1,\,\tilde{n}_2,
\cdots \tilde{n}_j\cdots$, where $\tilde{n}_m=\sum_s\,n_s$ for some
set, until the box is hit again. Since we are away from the
intermittent region we may expect that arrival probabilities in the
$M$ cells are essentially uniform, so, while the $\tilde{n}_l$
sequence has roughly a mean value which is $M$ times that of the $n_j$
sequence, we expect the tail of the distribution to be governed by the
same power law. Our numerical results suggest, once again, that, at
least asymptotically, {\it any} non-zero coupling 
induces instead uncorrelated behaviour on the two maps return
probabilities. Nevertheless the deviation from the uncoupled
distribution (thick-solid curve in Fig.~\ref{fig:5}(a)) takes place at
longer return-times as the coupling parameter $\ve$ is reduced
(thin-curves in Fig.~\ref{fig:5}(a)). The coupling parameter $\ve$
influences only the pre-asymptotic decay, suggesting the existence of
a scaling law: indeed observing Fig.~\ref{fig:6}(a) we find it by
writing $P(\tau)$ in the following form:  
\begin{equation}
  \label{Pscal}
  P_{rec}(\tau)\ \sim\ \ve^{\gamma\alpha}\mc{F}(\tau\ve^\alpha),
\end{equation}
where $\gamma$ is the power-law decay exponent, $\alpha$ is a fitting
parameter and $\mc{F}$ is a function independent on $\ve$. An
appropriate choice of $\alpha$ allows to get curve collapsing in the
asymptotic power-law regime. {We remark again that up to some
  $\ve$ depending transient time the dynamics is instead dominated by
  the uncoupled behavior.} In Fig.~\ref{fig:6}(b), the 
plot of $\alpha$ as a function of the intermittency parameter $z$
shows a behaviour that is quite accurately reproduced by a linear fit
with slope $\sim 4/5$. 

\subsection{Mixing test 2: Finite time Lyapunov exponents}
\begin{figure}[!htb]
  \centering
\includegraphics[width=\textwidth]{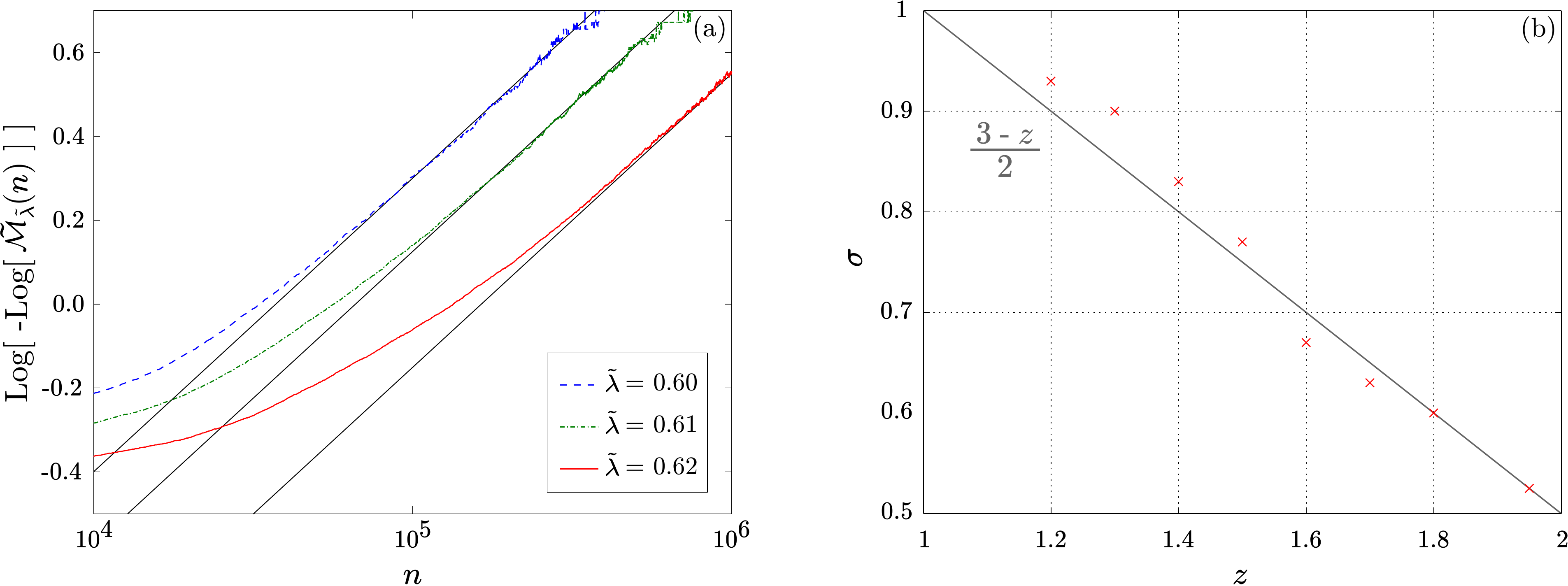}
\caption{(Color online) Panel (a): Examples of stretched exponential
  fits for the decay of 
  $\tilde{\mc{M}}_{\tilde{\lambda}}(n)\defeql{\mc{M}}_{\tilde{\lambda}}(n)/{\mc{M}}_{\tilde{\lambda}}(0)$
  ({\it i.e.} Eq.~(\ref{Mld}) normalized w.r.t. $n=0$ for  
  fitting purposes) for parameters $z=1.5$, $\ve=10^{-5}$ and three
  different values of $\tilde{\lambda}$; notice that the stretching
  exponent is the same for all three cases, but the exponential
  pre-factors are not. Panel (b): The stretching exponent $\sigma$ as a
  function of $z$ at $\ve=10^{-5}$; the straight continuous-black
  line is for reference.} 
  \label{fig:7}
\end{figure}

An alternative indirect way of probing mixing speed is provided by
large deviation analysis of the largest finite-time Lyapunov
exponents: this method is based upon rigorous results, proved in
Refs.~\cite{Mel,PoSh} (see also \cite{Luz}), and it was numerically tested
on intermittent and Hamiltonian systems in Ref.~\cite{am}. Since it is a
less widely used technique with respect to scrutinizing the statistics
of Poincar\'e recurrences, we briefly recall how it works
operationally: for further details see Ref.~\cite{am}. The crucial quantity
to take into account is the probability distribution of finite-time
Lyapunov exponents $\mc{P}_n(\lambda)$ (leading expansion rates up to
some fixed time $n$) which, for an ergodic system, collapses to a
Dirac delta in the $n \to \infty$ limit. The idea~\cite{Mel,PoSh,Luz}
is that, by fixing a threshold $\tilde{\lambda}<\lambda_\infty$
($\lambda_\infty$ being the asymptotic largest Lyapunov exponent) one
can define: 
\begin{equation}
  \label{Mld}
  \mc{M}_{\tilde{\lambda}}(n)=\int_0^{\tilde{\lambda}}\, d\lambda\, \mc{P}_n(\lambda),
\end{equation}
such that (in a large deviation philosophy) the way
$\mc{M}_{\tilde{\lambda}}(n)$ decays as $n \to \infty$ should also
rule the mixing speed. We underline that the important issue is
that intermittent dynamics deeply modifies large deviations
properties: the measure of point where Birkhoff sums up to time $n$
are at least $\epsilon$ different from the asymptotic average is
\textit{not} exponentially vanishing as $n\to \infty$, but decays
with a weaker power law, whose exponent coincide with the one ruling
correlations decay. The result is independent on the choice of 
threshold $\tilde{\lambda}$ as long as it is below the asymptotic
$\lambda_\infty$; however, in numerical implementations one has to
make preliminary checks (see Fig.~\ref{fig:7}(a)) to
fix the threshold reasonably (not too close to $\lambda_\infty$ to
spoil statistics, not too small to have only 
few points in the tail). The procedure we follow here (not knowing
{\it a priori} the invariant measure) is to follow a single
trajectory for $10^{12}$ time steps and re-construct
$\mc{P}_n(\lambda)|_{n\leq10^6}$ from $10^6$ consecutive chunks of
trajectory: the threshold is then fixed depending on the goodness of
the collected statistic. In Fig.~\ref{fig:7}(a) we find
that, like in the case of filling rates, the data are best fitted by a
stretched exponential decay $\mc{M}_{\tilde{\lambda}}(n)\sim
e^{-C\cdot n^\sigma}$; notice that different values of
$\tilde{\lambda}$ give the same stretching-exponent $\sigma$ but
different pre-factors $C$. We find that the exponent $\sigma$, at
diffusive coupling $\ve=10^{-5}$, display a roughly linear dependence
on parameter $z$ (Fig.~\ref{fig:7}(b)). Finally, we remark
that the discrepancy between the stretched exponential for the
$\mc{M}_{\tilde{\lambda}}(n)$ integral and the power-law decay for
Poincar\'e recurrences definitely deserves further studies, since, in  
principle, the two should be related. 

\newpage

\section{Conclusions}
\label{sec:conc}
In summary we have mainly investigated the ergodic properties of two
diffusively coupled, identical Pomeau-Manneville maps. 
In particular we have characterised how single trajectories fill fine
partitions of the phase-space: this yields an indication of
ergodicity, with a nontrivial filling rate in the form of a stretched
exponential. The same time-law appears in the decay of sub-threshold finite-time
Lyapunov exponent distribution, which probes the speed of mixing. By
contrast, a polynomial decay is instead observed when considering the
statistics of escape times and Poincar\'e recurrences. In addition,
the latter display a power-law exponent, for two coupled maps, which
is exactly twice the value for the single Pomeau-Manneville map,
suggesting some kind of asymptotic independence between the two
coordinates.  

\section*{Acknowledgements}
C.M. thanks CNPq, CAPES, FAPESC, Brazilian agencies, for financial
support. We are grateful to Bastien Fernandez for insightful
discussions on intermittent and diffusively coupled maps. 


\end{document}